\documentclass[a4paper,11pt]{article}
\pdfoutput=1 
\usepackage{tikz}
\usetikzlibrary{matrix}
\usepackage{amsmath} 

\usepackage{jcappub} 
\usepackage{bm}
\usepackage[T1]{fontenc} 
\usepackage{dirtytalk}
\usepackage{hyperref}

\def\x {\bm{x}}

\def\d {{\rm d}}

\def\k {\bm{k}}
\def\y {\bm{y}}

\def\aap{\ref@jnl{A\&A}} 
\def\bea{\begin{eqnarray}}
\def\eea{\end{eqnarray}}

\newcommand{\n}{{\hat{\bm n}}}

\newcommand{\three}{^{(3)}}
\newcommand{\two}{^{(2)}}
\newcommand{\one}{^{(1)}}

\newcommand{\<}{\big\langle}
\renewcommand{\>}{\big\rangle}

\newcommand{\red}[1]{{\color{red}{#1}}}

\newcommand{\average}[1]{\left\langle #1 \right\rangle}

\newcommand{\de}{\mathrm{d}}

 \title{\boldmath  {The effect of finite  halo size on the clustering of neutral hydrogen}}
 
\author[a]{Obinna Umeh,\footnote{Corresponding author.}}
\author[a,b]{Roy Maartens,}
\author[c]{ Hamsa Padmanabhan,}
\author[b,d,e]{and Stefano Camera}

\affiliation[a]{Institute of Cosmology \& Gravitation, University of Portsmouth, Portsmouth PO1 3FX,  United Kingdom.}
\affiliation[b]{Department of Physics, University of the Western Cape, Cape Town 7535,  South Africa.}
\affiliation[c]{D\'epartement de Physique Th\'eorique, Universit\'e de Gen\`eve 
24, quai Ernest-Ansermet, CH 1211 Gen\`eve 4, Switzerland}
\affiliation[d]{Dipartimento di Fisica, Universit\`a degli Studi di Torino, Via P.\ Giuria 1, 10125 Torino, Italy.}
\affiliation[e]{Istituto Nazionale di Fisica Nucleare -- INFN, Sezione di Torino, Via P.\ Giuria 1, 10125 Torino, Italy.}

\emailAdd{obinna.umeh@port.ac.uk}
\emailAdd{roy.maartens@gmail.com}
\emailAdd{hamsa.padmanabhan@unige.ch}
\emailAdd{stefano.camera@unito.it}

\abstract{{Post-reionisation 21cm intensity mapping experiments target the spectral line of neutral hydrogen (HI) resident in dark matter haloes. According to the halo model, these discrete haloes trace the continuous dark matter density field down to a certain scale, which is dependent on the halo physical size. The halo physical size defines an exclusion region which leaves imprints on the statistical properties of HI. We show how the effect of exclusion due to the finite halo size impacts the HI power spectrum, with the physical boundary of the host halo given by the splashback radius. Most importantly,  we show that the  white noise-like feature that appears in the zero-momentum limit of the power spectrum is \textit{exactly cancelled} when the  finite halo size is taken into consideration. This cancellation in fact applies to all tracers of dark matter density field, including galaxies. Furthermore, we show that the exclusion due to finite halo size leads to a sub-Poissonian noise signature on large scales, consistent with the results from N-body simulations. }
 
  }

\begin{document}
\maketitle
\flushbottom


\section{Introduction}




Cosmological information contained in the statistics of observed discrete sources (such as galaxies) is obtained by comparing the observations to a suitable theoretical model. There are several ways of building a theoretical model for each  of the discrete sources; the widely  used option involves treating these discrete sources as tracers of the stochastic non-linear dark matter density field.  As may be expected, this process of mapping discrete sources to dark matter involves a range of assumptions about the source in question. For example,  in the case of neutral hydrogen (HI) or 21cm intensity maps, the HI-bearing systems  are treated as residents of dark matter haloes, which, in turn, are treated as tracers of the stochastic non-linear dark matter density field \cite{Bull:2014rha,Santos:2015gra}. {In this framework, the dark matter haloes  are  extended virialized or gravitationally bound regions of space with density in excess of the cosmic  mean density \cite{Navarro:1995iw,Cooray:2002dia}.} Haloes come in different sizes and masses, and a specific  range of halo masses are physically connected to  HI-bearing systems  \cite{Santos:2015gra}. This halo mass range can very naturally be formulated in terms of circular velocity of the haloes \cite{Padmanabhan:2016fgy}, which also relates the HI content to the virial temperature and photoionization that suppresses the formation of dwarf galaxies \citep{rees1986, efstathiou1992}.  Modelling the statistics of HI as residents of dark matter haloes usually neglects how the finite size of the halo may impact the derived correlation function. We investigate this issue in detail in the present paper.

{It is well-understood in the context of halo statistics that 
due to the finite size of dark matter haloes, one must decide which structures are parent haloes and which  are sub-haloes of larger haloes. This choice  is known as the `halo exclusion criterion' \cite{CasasMiranda:2001ym,Sheth:2001dp,vandenBosch:2012nq,Baldauf:2013hka,Garcia:2019xel}}. Halo exclusion effects leave an imprint on the number density of haloes, their  correlation functions and halo bias parameters \cite{Garcia:2019xel}. It was pointed out during the early  development phase of the halo model of structure formation that halo exclusion must be taken into account for a more realistic halo model  to emerge \cite{Sheth:1998xe}.  Recent studies  within the halo model \cite{vandenBosch:2012nq, Garcia:2020mxz} have discussed several proposals on how to incorporate the halo exclusion effects in the modelling of halo correlation functions. A proposal on how to incorporate halo exclusion effects into standard perturbation theory  was given in \cite{Baldauf:2013hka}.  We build on the formalism discussed in \cite{Baldauf:2013hka} and apply it to the clustering of the  HI brightness temperature.  

We show how to model  HI host haloes of a given physical size or circular velocity.  The physical size is given by the splashback radius of the host halo \cite{Adhikari:2014lna,More:2015ufa}. We argue that the splashback radius provides a physical smoothing scale below which  the model of HI brightness temperature will need to take into account the effects of stellar streams and other {baryonic} physics at play within the halo \cite{Banik:2018pjp}.  Furthermore, we argue  that the splashback radius as a physical  length scale fits perfectly within the hierarchical cold dark matter structure formation picture. 
 The splashback radius is defined dynamically by the infall matter/particles on their first orbit \cite{Diemer:2014xya}, and it corresponds to the position of  sharpest drop in the slope of the dark matter density field \cite{Adhikari:2014lna,Diemer:2017uwt}.

Our analysis shows  that  the emergent white noise-like features that appear in the HI brightness temperature power spectrum in the limit  of zero momentum {(see  \cite{Umeh:2015gza,Umeh:2016thy,Penin:2017xyf})} 
 are \textit{exactly cancelled} when the finite size of the host halo  is taken into consideration.  
 {The white noise-like feature also arises in the galaxy power spectrum (see \cite{McDonald:2006mx,McDonald:2009dh,Assassi:2014fva})  and our argument applies in this as well.}
 In addition,  we show that there are sub-Poissonian noise signatures whose impact on the discrete HI power spectrum is naturally connected to the  mass-weighting of haloes introduced in \cite{Seljak:2009af}. 

The paper is organised as follows:  {In \autoref{sec:statistics}, we review the statistics of discrete sources and  describe the connection between the power spectrum of discrete sources and that of the continuous density field.  In \autoref{sec:HITb}, we describe how the HI brightness temperature may be modelled as a tracer of the dark matter density field within the halo framework. We compute and discuss the continuous power spectrum and show how the white noise-like term is cancelled exactly for tracers of finite size in \autoref{sec:finitesize0}.  A summary and conclusions are given in {sec:Discussionandconc}.} We provide basic tools for the halo model in \autoref{sec:halomodel}.
\\
{\bf{Cosmology}}: We adopt the following values for the cosmological parameters of the standard model \cite{Aghanim:2018eyx}: the dimensionless Hubble parameter, $h = 0.674$, baryon density parameter, $\Omega_{\rm b} = 0.0493$,  dark matter density parameter, $\Omega_{\rm{cdm}} = 0.264$,  matter density parameter, 
$\Omega_{\rm m} = \Omega_{\rm{cdm}} + \Omega_{\rm b}$,  spectral index, $n_{\rm s} = 0.9608$,  and the amplitude of the primordial perturbation, $A_{\rm s} = 2.198 \times 10^{-9}$.

\section{Statistics of discrete sources}\label{sec:statistics}

 \subsection{Probability of finding sources within a given volume}\label{sec:review}
 
 The probability of finding two {discrete tracers of type $X$}, in small volumes $\delta V_{1}$ and $\delta V_{2}$, and separated by a distance ${\x}_{12}$, is given by {(omitting redshift dependence for brevity)}
 \begin{equation}\label{eq:def_probability}
 \delta P_{12}({\x}_{12}) = \bar{n}^2_{X} \left[ 1 + \xi_{X} ({\x}_{12}) \right] \delta V_{1}\delta V_{2}\,,
 \end{equation}
where  {$\bar{n}_{X}$ is the average number density and} $\xi_{X}({\x}_{12})$ is the two-point correlation function (2PCF) that describes the excess probability, compared to random, of finding sources separated by ${\x}_{12}$. In the isotropic limit, $\xi({\x}_{12}) = \xi(|\x_{12}|)$ is independent of direction {and orientation of the pair}. $\xi_{X}$ is subject to the following conditions due to the physical meaning of probability:
 \begin{enumerate}
 \item $\xi_{X}({\x}_{12} )\ge -1$, since probability must be {non-negative}; saturation of the bound, i.e.\ zero probability, corresponds to the exclusion limit \cite{Sheth:1998xe}.
 \item {$\xi_{X}({\x}_{12}) \rightarrow 0$ as ${|{\x}_{12}| \to0}$}  is required for $\bar{n}_{X}$ to correspond to an observable mean number density. 
 \end{enumerate}

In practice, the 2PCF is estimated by counting the number of source pairs {within} volumes in a given catalogue, and comparing it to the number that would be expected on the basis of a Poisson distributed catalogue with the same total population. These catalogues are generated with some level of arbitrariness, for example:
\begin{itemize}
\item {\bf{Halo catalogue:}} Every halo catalogue depends on the  halo exclusion  length scale.  This is a selection effect associated with the criteria adopted for assigning structures as parent haloes and sub-haloes of the larger halo.  For instance,  two structures with  masses $M_1$ and $M_2$, are considered to correspond to the same parent halo if the separation between them, $x_{12}$, satisfies \cite{Garcia:2019xel}
\begin{equation}\label{eq:exclusioncriteria}
x_{12}\equiv |{\x}_{1}- {\x}_{2}| \le R(M_1,M_2)\,,
\end{equation}
where $R$ is a characteristic scale, whose choice has an impact on the halo statistics \cite{Garcia:2019xel}. 
There are several choices for $R$, but the key point is that $x_{12}$ is finite for massive haloes. Various criteria were considered in \cite{Garcia:2019xel}, which found that each choice leaves its corresponding imprint on halo statistics such as the halo mass function, correlation function and  clustering bias.
 
  \item {\bf{Galaxy catalogues}}:  Although we focus on the consequences of the host halo finite size on the correlation functions of the HI brightness temperature, we note that the same technique can also be applied to galaxy correlation functions;  the difference is in the treatment of number-weighted halo occupations for galaxies, in contrast to mass-weighted ones in the case of intensity mapping.   Most mock galaxy catalogues start from dark matter  haloes produced from a given $N$-body dark matter simulation \cite{Carretero:2014ltj}, which serve as the locations to place galaxies. The specific ingredients used to populate the dark matter haloes differ from model to model. 

 \item {\bf{HI intensity map  catalogues}}: 
 Catalogues of HI intensity maps are produced by populating the halo catalogue with recipes  that describe the distribution of HI atoms within the halo. This distribution is determined by the HI-halo mass relation,
 \begin{equation}\label{eq:HImassfunction}
 M_{\rm HI}(R_{c} )= \int_{0}^{R_{c}} 4\pi r^2{\varrho}_{\rm HI} (r)\, \de r \,,
 \end{equation}
 where ${\varrho}_{\rm HI}$ is the HI radial profile {(assuming spherical symmetry)} and $R_{c}$ specifies the boundary of the HI-bearing dark matter halo in the limit of spherical symmetry.
 In this way, we relate the exclusion region to the HI brightness temperature.   
 
Considering the astrophysics of the HI brightness temperature, we see that $R_{c}$  is related to the circular velocity of the host halo as
\bea
 R_{c}^2 = {G {M} \over v^2_c}, 
 \label{defrc}
 \eea
 where ${M}$ is the mass  of the dark matter halo that can host HI. The fitting function that describes the dependence of  $M_{\rm HI} $  on $v_c$ and ${M}$ is  given by \cite{Padmanabhan:2016fgy,Padmanabhan:2018llf}
\begin{equation}\label{eq:MImass}
M_{\rm HI}(v_{c0},{M}) =  \phi_{\star}({M}) \left(\frac{{M}}{M_{\star}}\right)^{\beta} 
\exp\left[-\left(\frac{v_{{\rm c0}}}{v_{\rm c}({M})}\right)^3\right]\,,
\end{equation}
where $M_{\star} =10^{11}\,h^{-1}M_{\odot}$,   $ \phi_{\star}({M}) =  \alpha f_{\rm H,c} {M}$, and  $v_{c0} = 36$ km/s. The term $v_{c0}$ is interpreted as the minimum circular velocity a halo requires to be able to host HI. The remaining parameters are: $\alpha=  0.09$; the average HI fraction relative to the cosmic fraction $f_{\rm H,c} = (1-Y_{p}) \Omega_{\rm b}/\Omega_{\rm m}$, where $Y_p = 0.24$ is the cosmological helium fraction \cite{Pitrou:2020etk}; and $\beta = -0.58 $, the logarithmic slope of 
the HI-halo mass relation. The circular velocity as function of ${M}$ is given by \cite{Camera:2019iwy}
\begin{equation}
    v_c ({M})= 96.6 \,\mathrm{km/s} \left(\frac{\Delta_c \Omega_{\rm m} h^2}{24.4} \right)^{1/6} \left(\frac{{M}}{10^{11} M_{\odot}}\right)^{1/3} \left(\frac{1+z}{3.3}\right)^{1/2}\,,
\end{equation}
where $\Delta_c=18\pi^2+82x-39x^2$ \citep{bryan1998},
with $x=\Omega_{\rm m}(1+z)^3/[\Omega_{\rm m}(1+z)^3+\Omega_\Lambda]-1$. 
In this case, one could infer that the the exclusion scale is determined by the minimum circular velocity of haloes that support HI. 
We explore this connection in more detail in \autoref{sec:HITb2}. 



\end{itemize}

\subsection{Connection between discrete sources and continuous  field }\label{sec:linktocontinuousfield}

The standard  practice in cosmology is to treat sourcedensity contrasts as peaks, a population of objects described by the density field of a point process. In this limit, the density contrast of a population of discrete sources $X$ is given by \cite{Bardeen:1985tr,Cooray:2002dia}
 \begin{equation}\label{eq:overdensity}
 \delta^{(d)}_{X} ({\x})\equiv \frac{n_{X}({\x})}{\bar{n}_{X}} - 1 = \frac{1}{\bar{n}_{X}} \sum_{i} \delta^{({\rm D})} ( {\x} - {\x}_i) -1\,,
 \end{equation}
 where  $n_{X}$ is the  comoving number density, ${\bar{n}}_{X}$ is defined in \autoref{eq:def_probability} and $ \delta^{({\rm D})} $ is the Dirac delta function. 
The  2PCF of the source density contrast then is defined as
 \begin{equation}
 \xi^{(d)}_{X}({\x}_1,{\x}_2) \equiv \< \delta^{(d)}_{X}({\x}_1)\,\delta^{(d)}_{X}({\x}_2)\> = \frac{1}{\bar{n}^2_X} \< n_{X}({\x}_1)\,n_{X}({\x}_2)\> -1\,,
 \end{equation}
{with angle brackets denoting ensemble averages.}

From \autoref{eq:overdensity},  $ \< n_{X}({\x}_1)n_{X}({\x}_2)\>$  can be decomposed into two parts: the part describing correlation when the two points are the same, and the part describing correlation when the points are different, i.e.\
 \begin{equation}\label{eq:decomposition}
 \< n_{X}({\x}_1)\,n_{X}({\x}_2)\>= \<\sum_{i} \delta^{({\rm D})}({\x}_1 - {\x}_i)\, \delta^{({\rm D})}({\x}_2-{\x}_i)\> + \sum_{ij}  \<\delta^{({\rm D})} ({\x}_1 - {\x}_i)\, \delta^{({\rm D})}({\x}_2-{\x}_j)\>\,. \qquad
 \end{equation}
Using \autoref{eq:decomposition}, the  discrete 2PCF becomes 
\begin{equation}\label{eq:obstocont}
  \xi_{X}^{(d)}({\x}_1,{\x}_2) =  \xi_{X}^{(c)}({\x}_1 -{\x}_2) + \frac{1}{\bar{n}_{X} }\delta^{({\rm D})}({\x}_1-{\x}_2)\,.
 \end{equation}
 The discrete 2PCF is made up of a continuous  2PCF $\xi_{X}^{(c)}$ and  the shot noise due to a Poisson process, which is given exclusively by the mean number density of the tracer $X$.
The relevant Fourier transforms are given by 
\begin{align}\label{eq:2PCFtransform}
\xi_{X}({\x}_1 -{\x}_2)& = \int
\frac{\d^3 \k}{(2\pi)^3}   \,{\rm e}^{i{\k}\cdot ( {\x}_{1} - \x_2)} 
P_{X}({\k}) \,, \\
 \delta^{({\rm D})}({\x}_1-{\x}_2) &=  \int
 \frac{\d^3 \k}{(2\pi)^3}\, {\rm e}^{i{\k}\cdot ( {\x}_{1} - {\x}_2)}\,,
 \label{eq:deltafunction}
\end{align}
where $P_{X}$ is the  power spectrum.
Using  \autoref{eq:2PCFtransform}  and  \autoref{eq:deltafunction}, \autoref{eq:obstocont} becomes \cite{Peebles:1976A}
 \begin{equation}\label{eq:Peeblesapproximation}
 P^{(d)}_{X}({\k}) \equiv P_{X}^{(c)}({\k}) +  \frac{1}{\bar{n}_{X}}\,.
 \end{equation}
 
In deriving \autoref{eq:Peeblesapproximation}, we have assumed through the use of \autoref{eq:overdensity}, \autoref{eq:2PCFtransform}  and  \autoref{eq:deltafunction}  that the discrete sources are correlated in all regions of space.  This is not strictly true when one considers  the geometry  of the  discrete tracers.  Discrete sources have well-known and discernible boundaries \cite{Garcia:2020mxz}.  
To model the geometry of sources we observe closely enough, we decompose the continuous 2PCF in \autoref{eq:obstocont}  further:
 \begin{equation}
    \xi_{X}^{(c)}(|{\x}_1 -{\x}_2|)  =    \xi_{X}^{(c)}(|{\x}_1 -{\x}_2| <R)  +    \xi_{X}^{(c)}(|{\x}_1 -{\x}_2| \ge R) \,,
 \end{equation}
where {we have imposed isotropy and}  $R$ is the comoving length scale which is associated with the  size of the halo in the case of HI brightness temperature. Shortly, we shall describe  how it is related to $R_c$ introduced in \autoref{eq:HImassfunction}.  We assume that the physics responsible for tracer clustering on scales $x_{12} >R$ is uncorrelated with the physics responsible for the dynamics on scales $x_{12} \le R$. In the effective field theory language, we integrate out modes with wavelength less than $R$. 
This implies that $\xi_{X}^{(c)}(|{\x}_1 -{\x}_2| <R) = -1$, where $|{\x}_1 -{\x}_2| <R$ is known as the exclusion region \cite{Sheth:1998xe}. 

Hence, the full decomposition of  $  \xi_{X}^{(d)}$ becomes
 \begin{equation}\label{eq:nonpertcondition}
 \xi^{(d)}_{X}({|{\x}_1 -{\x}_2|}) {=}
\begin{cases}
{\bar{n}_{X}^{-1}}\qquad \qquad \qquad ~~~ {\rm{for}} \quad & |{\x}_1 -{\x}_2| = 0 \,, \\
   -1 \qquad \qquad \qquad ~~~~ {\rm{for}} \quad &0< |{\x}_1 -{\x}_2| < R\,,\\\ 
   \xi_{X}^{(c)}({|{\x}_1 -{\x}_2|}) \quad~~~ {\rm{for}} \,\,& |{\x}_1 -{\x}_2|  \geq R\,. \\
 \end{cases}
 \end{equation}
The last term is the part of the 2PCF which can be modelled as a  continuous {field} on scales $|{\x}_1 -{\x}_2|  \geq R$. 
Taking the inverse Fourier transform of $ \xi^{(d)}_{X}({|{\x}_1 -{\x}_2|})$, but this time accounting for the condition given in \autoref{eq:nonpertcondition}, leads to 
 \begin{equation}
 P^{(d)}_{X}(k) 
   =\frac{1}{\bar{n}_{X}} - \int_{{x_{12}<R}} \d^3 \x_{12}\,{\rm e}^{-i{\k}\cdot {\x}_{12}} + \int_{{x_{12}\geq R}} \d^3 \x_{12}\, \xi_{X}^{(c)} (x_{12})\,{\rm e}^{-i{\k}\cdot {\x}_{12}} \,.
 \end{equation}
 The second term { gives the standard Fourier transform of a top-hat window:
 \bea \label{vrwr1}
\int_{{x_{12}<R}} \d^3 \x_{12}\,{\rm e}^{-i{\k}\cdot {\x}_{12}} =4\pi\int_0^R \d r\, r^2\,j_0(kr) = V_R W_R(k)\,,
\eea
where 
 \bea\label{eq:VolandW}
 V_{{R}} &=& \frac{4\pi}{3} R^3 \,,\\ \label{vrwr2}
 W_{R}(k)  &= &3\left[ \frac{ \sin(kR) - kR\cos(kR)}{(kR)^3}\right] = {3 \,{j_1(kR) \over kR}}  \,.
 \eea
 Here, $j_1$ is a spherical Bessel function  of order one and $ V_{{R}}$ is the excluded volume modulated by the window function $ W_{R}(k)$.
 Then we have
 \begin{equation}
 P^{(d)}_{X}(k)   =\frac{1}{\bar{n}_{X}} - V_{{R}} W_{R}(k) -\int_{{ x_{12}< R}} \d^3 \x_{12}\,\xi_{X}^{(c)} (x_{12}) \,{\rm e}^{-i{\k}\cdot{\x}_{12}} 
   + {\int} \d^3 \x_{12}\,\xi_{X}^{(c)} (x_{12})\,{\rm e}^{-i{\k}\cdot {\x}_{12}} \,.
    \end{equation}
 Performing the inverse Fourier transform for the remaining terms, we find that
 \bea\label{eq:fullpower}
   P_{X}^{(d)}(k) =\frac{1}{\bar{n}_{X}} -V_{{R}} W_{R}(k) - \big[W_{R}\star P_{X}^{(c)}\big]({k}) 
   + P_{X}^{(c)}(k)\,,
 \eea
where $W_{R}\star P_{X}^{(c)}$ is a convolution of the exclusion {window} and the continuous power spectrum. We show how to perform the convolution integral exactly in \autoref{sec:finitesize}.

\section{ Clustering of HI brightness temperature in perturbation theory }\label{sec:HITb}

HI intensity mapping is an observational technique for mapping the large-scale structure of the Universe in three dimensions using the integrated {21cm} emission from gas clouds, without the requirement to resolve individual galaxies.  
The 21cm emission line arises from the spin-flip transition in hydrogen atoms, and is a unique probe of the hydrogen density at a particular frequency, {allowing intensity mapping surveys to answer fundamental questions on the origin and evolution of large-scale cosmic structures}.

In the Rayleigh limit, the HI intensity is related to its brightness temperature   
\begin{equation}\label{eq:HIbrightnesspert}
T_{{\rm HI}} (z, {\n})= \frac{3 h c^3 A_{10}}{32 \pi k_{B} \nu_{21}^2} \frac{(1+z)^2}{H(z) }n_{\rm HI} (z, {\n}) = C_{\rm HI}(z) \bar{n}_{\rm HI}(z) \left[1 + \delta_{\rm HI}(z, {\n})\right]\,,
\end{equation}
where $z$ is the redshift and  ${\n}$ is the  line of sight direction of the source. The number density of HI atoms  is expanded perturbatively as $n_{\rm HI} = \bar{n}_{\rm HI} (1 + \delta_{\rm HI})$, introducing the mean HI number density, $\bar{n}_{\rm HI}$, and {fractional} 
perturbation, $\delta_{\rm HI}$.  $C_{\rm HI}$ is an {amplitude depending on} physical constants and background cosmology parameters \cite{Hall:2012wd}:
\begin{equation}
C_{\rm HI}(z) = \frac{3 h c^3 A_{10}}{32 \pi k_{\rm B} \nu_{21}^2} \frac{(1+z)^2}{H(z) }\,,
\end{equation}
where $H$ is the Hubble rate, $\nu_{21}$ is the {rest-frame} frequency of emitted photons, and $A_{10}= 2.869 \times 10^{-15}\, {\rm s}^{-1}$ is the emission rate.

Our modelling of $\delta_{\rm HI}$ relies on perturbation theory, since  the  evolution of structures  become highly non-linear and even non-perturbative on small scales. We go beyond linear order, to account for the non-linear effects, by including the one-loop corrections to the power spectrum \cite{Umeh:2016thy}. On non-perturbative scales, HI can be `painted' on to dark matter  in $N$-body  simulations by using prescriptions such as the one  in \autoref{sec:linktocontinuousfield} \cite{Seehars:2015ada}.  Hydrodynamical simulations can also be used  to model the distribution of HI  \cite{Villaescusa-Navarro:2018vsg}. 


\subsection{Smoothing of high-frequency modes}\label{sec:HITb2}


The HI fluctuations $\delta_{\rm HI}$ introduced in \autoref{eq:HIbrightnesspert} are given by 
\begin{equation}\label{eq:definition}
\delta_{\rm HI}  (z,{\x}) \equiv \frac{{n _{\rm HI}(z,{\x}) -\bar{n}_{\rm HI} (z) }}{\bar{n}_{\rm HI}(z) } \,.
\end{equation}
This implies that, by definition, the volume average of $\delta_{\rm HI}$ vanishes, $\<\delta_{\rm HI} (z,{\x})\> = 0$, since the mean number density is defined as  $\bar{n}_{\rm HI} (z)  \equiv \<n _{\rm HI}(z,{\x}) \>$.
 However, this {condition} is violated when a bias model is used to relate $\delta_{\rm HI}$ to the underlying dark matter density field, with fractional perturbation $\delta_{\rm m}$.  For example, consider a simple Eulerian bias model, where $\delta_{\rm HI}$ is only a functional of the local matter density,  i.e.\ ${\delta_{\rm HI} = F[\delta_{\rm m}]}$ \cite{Umeh:2015gza,Desjacques:2016bnm}. In this case,
\begin{equation}\label{eq:n_m_expansion}
   n_{\rm HI}(z,{\x}) = \bar{n}_{\rm HI}(z)\left[1 + b_1(z)\delta_{\rm m}(z,{\x}) 
    + \frac{1}{2!}b_{2}(z)\delta^2_{\rm m}(z,{\x})
    +\frac{1}{3!}b_{3}(z)\delta^3_{\rm m}(z,{\x})+ \mathcal{O}(\delta_{\rm m}^4)\right],
\end{equation}
where  $b_{i}$ are the {$i$th-order} HI bias parameters. (For simplicity, we neglect tidal and derivative bias parameters.) 
Taking the spatial average of \autoref{eq:n_m_expansion} leads to
\begin{equation}\label{eq:nonvanishing}
    \average{n_{\rm HI}(z,{\x}) }= \bar{n}_{\rm HI}(z)\left[1 
    + \frac{1}{2!}b_{2}(z)\average{\delta_{\rm m}^2(z,{\x})}+ \frac{1}{3!}b_{3}(z)
    \average{\delta^3_{\rm m}(z,{\x})}
+ \mathcal{O}(\delta_{\rm m}^4)\right],
\end{equation}
with $\average{\delta_{\rm m}^2(z,{\x})} $ being  the variance of the dark matter density field, $\sigma_{\rm m}^2$, and $\average{\delta^3_{\rm m}(z,{\x})} $ is the skewness, $S_{3}$.  The skewness vanishes in the  Gaussian limit, which we henceforth focus on. 

One way to  ensure that the spatial average of $\delta_{\rm HI}$ vanishes, is to subtract $\sigma_{\rm m}^2{(z)}\equiv \average{\delta_{\rm m}^2({z},{\x})} $ from both sides of \autoref{eq:n_m_expansion}, so that
\begin{equation}\label{eq:renormalization}
\delta^2_{\rm m}(z,{\x})\to  \delta^2_{\rm m}(z,{\x}) - \sigma_{\rm m}^2(z)\,.
\end{equation}
As a consequence, the mean number density changes as ${ \bar{n}_{\rm HI}\to \bar{n}_{\rm HI} (1 +\sigma_{\rm m}^2/2)}$.  This process has some issues because  $\sigma_{\rm m}^2$ does not behave well in the {non-perturbative regime $k\gg k_{\rm NP}$}, calling into question the validity of the  perturbative expansion. There are three ways that this may be handled. 

First, one could introduce an arbitrary hard ultraviolet cut-off \cite{Umeh:2015gza,Umeh:2016thy,McDonald:2006mx,McDonald:2009dh}, but this will mean that the re-defined mean number density, bias parameters, and other physical quantities depend upon the arbitrary cut-off. A second option is to adopt the effective field theory (EFT) approach and introduce an EFT scale, $\Lambda$, such that modes with  $k>1/\Lambda$ are integrated out \cite{Assassi:2014fva,Desjacques:2016bnm}, and the bias parameters are consequently rewritten as `renormalised bias parameters' in order to suppress their dependence on the EFT scale. The third option   is to introduce a smooth physically motivated cut-off, which naturally describes the geometry of the exclusion region. This is the option  we adopt here. We connect it  with evidence to show that the smoothing scale is determined by the physical size of the host halo \cite{Wadekar:2020oov}.

In Fourier space, we model {the smooth cut-off with a window function $W_{R}(k)$, which suppresses the contribution from $k> 1/R$}:
\begin{equation}\label{eq:WindowHI}
 {\delta_R(z,\k) \equiv} \delta_{\rm m}^{R}({z},{\k})  = W_{R}(k) \delta_{\rm m}({z},{\k})\,.
\end{equation}
Here $R$ is not an arbitrary scale. This approach differs in principle from the model described in \cite{Schmidt:2012ys}, where $R$ is the radius of an arbitrary averaging domain.  
 In real space, \autoref{eq:WindowHI} leads to a convolution
\begin{equation}\label{eq:densityfilter}
 { \delta_{R}}(z,{\x}) =\int \d^3 {\bm{y}}\,W_{R}(|{\x}-{\y}|)\delta_{\rm m}(z,{\y})\,,
\end{equation}
where $W_{R}$ specifies the physical boundary, and we use a top-hat filter, given the result in \autoref{eq:VolandW}. By convolving the dark matter density field with a top-hat filter function in real space, \autoref{eq:densityfilter} helps  to parametrically filter out the high-frequency modes that we are not sensitive to in the dark matter density field in Fourier space \cite{Schmidt:2012ys,Desjacques:2011mq}. {This removes} the bad ultraviolet behaviour in the dark matter variance:
 \begin{equation}
 \sigma_{R}^2(z)=\left<\delta^2_{R}(z,{\x})\right> = \frac{1}{2\pi^2} \int \d k \,k^2 W_{R}^2(k) P_{\rm m}(z,k) \,,
 \end{equation}
  where $P_{\rm m}$ is the matter power spectrum.
  
In this case, the  re-defined mean HI number density becomes
\begin{equation}\label{eq:reno2}
\bar{n}^R_{\rm HI}(z) 
={ \bar{n}_{\rm HI}(z) \Big[1 +  \frac{1}{2}\sigma_{R}^2(z)\Big]}\,,
\end{equation}
so that the  HI density fluctuation becomes\footnote{Note that this bias relation holds provided that the  wavelengths of the dark matter density {modes are} larger than the size of the host {halo}.} 
\begin{equation}\label{eq: n_m_expansion2}
\delta_{\rm HI}^{R} (z,{\x}) 
  =  b^{R}_1(z)\delta_R(z,{\x}) 
    + \frac{1}{2}b^{R}_{2}(z)\left[\delta_R^2(z,{\x})-\<\delta_R^2\>{(z)}\right]
  +\frac{1}{3!}b^{R}_{3}(z){\delta_R^3(z,{\x})}+  { \mathcal{O}(\delta_{R}^4)}\,,
\end{equation}
where we have re-defined the HI bias parameters as
\begin{equation}\label{eq:redefinedbiasparameters}
b_{i}^R(z)={b_{i} (z)\over 1 +   {\sigma_{R}^2(z)/2}}.
\end{equation}
The re-defined HI density fluctuation now averages  to zero, $\average{\delta_{\rm HI}^{R} (z,{\x})} = 0$,  restoring the consistency of perturbation theory.  {The renormalised bias parameters are shown in \autoref{fig:biasR}. 
  \begin{figure}[t!]
  \centering
\includegraphics[width=70mm,height=65mm ]{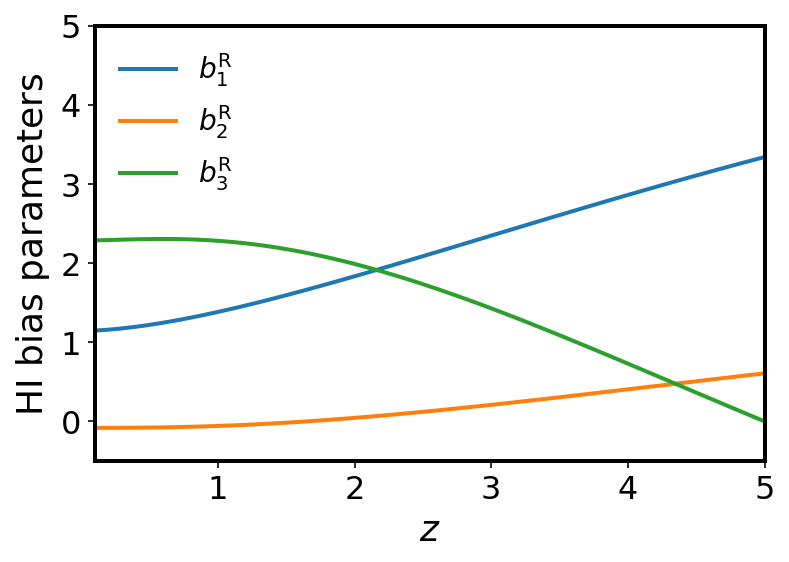}
\includegraphics[width=70mm,height=65mm ]{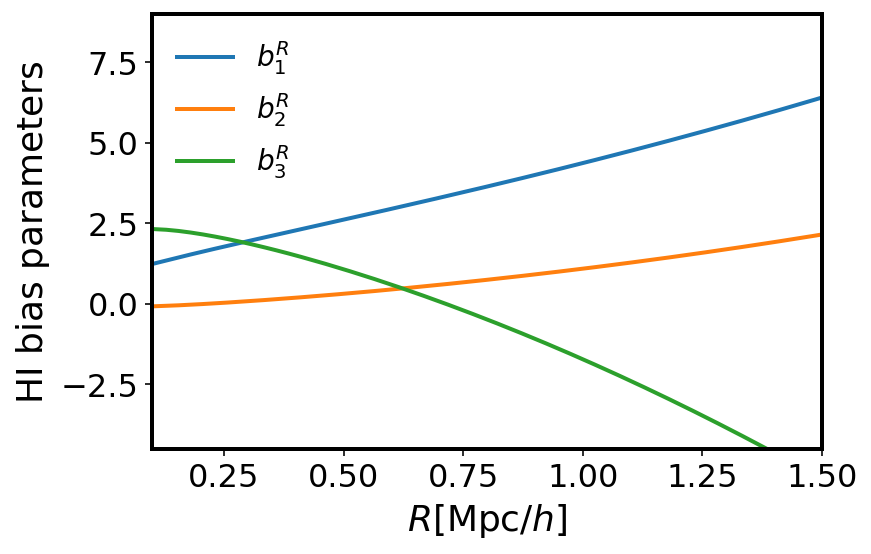}
\caption{Re-defined HI bias parameters in \autoref{eq:redefinedbiasparameters} as a function of redshift (left), {and re-defined HI bias parameters at $z=1$ as a function of $R$ (right).} 
}
\label{fig:biasR}
\end{figure}

The validity of the perturbation theory expansion requires that we  restrict attention to scales where  the standard deviation is less than unity. 
For the HI density contrast  this implies that $|b^{R}_1 \sigma_{R}(z)| < 1$. On small scales, $R <6 \,{\rm Mpc}/h$ at $z =0$, the standard deviation  of the dark matter density field is $\sigma_{R}(z) >1$.  However, as described in \cite{Schmidt:2012ys}, the necessary condition for convergence is that $b_{n+1}^R/b_{n}^R$ is approximately constant at order $n$. The trade-off is that one needs to go to  higher order in perturbation theory to get a converged expression. 
%

One last important feature to note is  that $\bar{n}^{R}_{\rm HI}$, $\delta_{\rm HI}^{{R}}$, and $b_{i}^R$   are now dependent on the size of the domain containing HI:
\begin{align}
\frac{\partial \bar{n}^{R}_{\rm HI}(z)}{\partial R}\,,~ 
 \frac{\partial \delta_{\rm HI}^{R}(z,{\x})}{\partial R}\,,~ 
  \frac{\partial b_{i}^R(z)}{\partial R} \neq 0\,. 
\end{align}
This is in agreement with the findings from the analysis of N-body simulations, which shows that these parameters are dependent on the exclusion scale which is determined by the halo mass~\cite{Garcia:2019xel}. 

\subsection{Dependence of HI statistics on splashback radius}

We calculate the HI bias parameters defined \autoref{eq:redefinedbiasparameters} from the model of the local density contrast (halo model) by weighting the halo bias parameters with the HI-halo mass relation: 
\begin{equation}\label{eq:bias_parameters}
   b^R_{i}(z)  = \frac{1}{{\bar\rho}_{\rm HI}(z) } \int_{M_{\rm{min}}(z)}^{\infty} \de M\, b^i_{h}(z,M)\,M_{\rm HI}  (v_{c0},M)\, n_h(z,M)\,,
\end{equation}
where ${M}$ is the halo mass, 
$n_{h}$  is the halo mass function,  $ b^i_{h}$ are the {$i$th-order} halo bias parameters,\footnote{ The full expressions for $n_{h}$ and $b_{h}$  are given in Appendix  \ref{sec:halomodel}, using the standard Sheth-Tormen halo mass function \cite{Sheth:1999mn}.   } and  ${\bar\rho}_{\rm HI}$ is the mean {comoving} density of HI, 
defined as the first moment of the HI-halo mass  function:
  \begin{equation}
{\bar\rho}_{\rm HI}(z) = \int_{M_{\rm{min}}(z)}^{\infty} \de M\,
 M_{\rm HI}(v_{c0},M)\,n_h(z,M)\,.
\end{equation}
Here, $M_{\rm{min}}$ is the 
minimum  mass a halo must have in order to  host HI.

We need to describe the relationship between the mass enclosed by the comoving sphere of radius $R$, and $M_{{\rm{min}}}$ or $v_{c0}$ introduced in \autoref{eq:bias_parameters}. We start  with the definition of the halo mass introduced in equation \autoref{eq:bias_parameters}.  The halo mass $M$ is defined with respect to a spherical top-hat filter as mass contained within a region of space with density contrast greater than the critical density of the universe by a factor $\Delta_c$ \cite{Sheth:1999mn}
\begin{equation}\label{eq:halomass}
 M \equiv  \frac{4\pi}{3}\bar \rho_{\rm{c}}(z) \Delta_c(z) R_{c}^3(z)  = {M_{c}}(z)  \,,
\end{equation}
where ${\bar\rho}_c$ is the  critical density of the universe at redshift $z$.
\begin{figure}[h!]
  \centering
\includegraphics[width=100mm,height=80mm ]{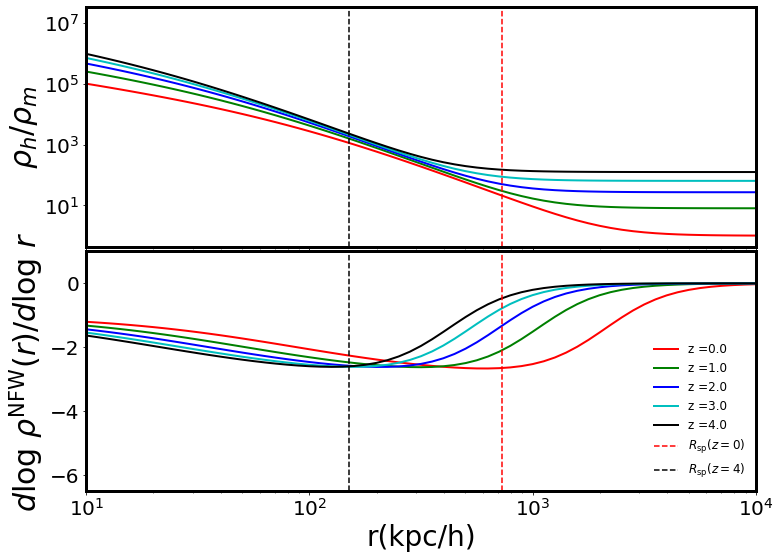}
\caption{ {Navarro-Frenk-White halo density  profile (top) and its slope (bottom) for a range of redshifts. Splashback radius is shown by vertical lines (black at $z=4$, red at $z=0$).  
}
 }
\label{fig:halo_profile}
\end{figure}

The radius, $R_{c}$, does not correspond to the physical boundary of the halo because the mass contained within $R_{c}$ is subject to pseudo-evolution {(i.e., evolution of the halo mass due to the evolution of the density at the reference redshift \cite{Diemer:2012mw})},   which breaks mass conservation~\cite{More:2015ufa}.  In addition, recent studies have shown that  sub-structures which form during collapse involve processes that redistribute mass from small to large radii greater than $R_{c}$~\cite{Kazantzidis:2005su,Valluri:2006vi,Carucci:2014ema}.  
 For these reasons, in \cite{More:2015ufa}  a coordinate-independent definition of the halo boundary was introduced, which includes all matter that orbits the main halo -- known as the `splashback radius', $R_{\rm sp}$\footnote{{Note that the natural physical scale for HI in the present context is given by the halo virial velocity cutoff~\cite{Padmanabhan:2014zma} or the splashback radius scale that we use here, since HI here traces collapsed, virialized dark matter haloes (rather than, for example, the  Jeans length which describes HI clumps in the Lyman-alpha forest ~\cite{Schaye:2001hu}, which are low column density systems outside collapsed structures).}}. It is dynamically defined by particles that reach the apocentre of their first orbit after infall \cite{Diemer:2014xya,Adhikari:2014lna}. There is a  pile up of particles at the apocentre due to their low radial velocity, thereby  creating a caustic that manifests as a sharp drop in the density profile in the halo outskirts , as shown in \autoref{fig:halo_profile}. 
It  has been detected in the Sunyaev-Zel'dovich signal of galaxy clusters \cite{Aung:2020grs,Shin:2018pic}  and in 3000 optically selected galaxy clusters over a redshift range  $0.1 <  z< 1.0$  from the Hyper Suprime-Cam Subaru Strategic Program \cite{Murata:2020enz}. Improvements of the theoretical modelling of clustering given the splashback radius  are currently being pursued~\cite{Diemer:2017uwt,Adhikari:2018izo,Diemer:2020xcz,Kazantzidis:2005su,Valluri:2006vi,Carucci:2014ema}\footnote{See \url{http://www.benediktdiemer.com/research/splashback} for an exhaustive list of related efforts.} .  



%
%

To connect $R_{\rm sp}$ to $M_{{\rm{min}}}$,  wel use the fitting function given in \cite{More:2015ufa}, which specifies the relation between $R_{\rm{sp}}$ and $R_c$, then use \autoref{eq:halomass} to relate $R_{\rm sp}$ to $M_{{\rm{min}}}$:
\begin{equation}\label{eq:Radiusratio}
\frac{R_{\rm{sp}}{(z)} }{R_{c}(z)} = A(z) + B(z)\, {\rm e}^{{\Gamma(z)}/C(z)}\,, 
\end{equation}
where $\Gamma $ is the mass accretion rate and $A$, $B$, $C$ are given in \cite{More:2015ufa} as
{
\begin{align}
A(z) &= 0.54\big[1 + 0.53\,{\Omega_{\rm m}(1+z)^3H_0^2/H^2(z)}\big]\,,
\\
B(z)&= {1.36\,A(z)\,,}
\\
C(z)&= 3.04\,.
\end{align}
}
Although these  fitting functions where  obtained in  \cite{More:2015ufa} at  fixed $\Gamma $, subsequent studies have shown that they evolve with redshift \cite{Diemer:2017uwt}. This is most likely due to the physics of mass accretion, which dominates mass growth at very high redshift \cite{Cole:2000ex,Behroozi:2014tna,Becker:2015iqa,ODonnell:2020dqv}. 
Thus, we parametrise $\Gamma $ as $\Gamma(z) = \Gamma_{1} + \Gamma_{2} z$.  
%
%
Using \autoref{eq:Radiusratio} and \autoref{eq:halomass},  we find that  $M_{\rm min}$ can be expressed in terms of the splashback radius as
\begin{equation}\label{eq:ModelMmin}
M_{\rm{min}}(z|\theta ) =  \frac{4\pi}{3}\bar \rho_{\rm{c}}(z) \Delta_c(z) \left[\frac{R_{\rm{sp}}{(z)}}{ A(z) + B(z) \,{\rm e}^{-\Gamma/C(z)}}\right]^3\,,
\end{equation}
where $\theta \equiv  \left\{ R, \Gamma_{1}, \Gamma_{2}\right\}$ are physical parameters.
Note that $R_{\rm{sp}} $ is a physical (i.e.\ proper) radius \cite{Adhikari:2018izo}. The corresponding comoving scale  is related to $R_{\rm{sp}} $ according to $R_{\rm{sp}} =  {R /(1+z)}$. 
We fix the parameter values by comparing our model of the HI density parameter (which represents the comoving density fraction of HI), to the corresponding measurements of $\Omega_{\rm HI}$ made at various redshifts~ \cite{Hamsa2015}. 
Within the halo model,  $\Omega_{\rm HI}$ is given by
\begin{equation}
\Omega_{\rm HI}(z|\theta)\equiv 
\frac{{\bar\rho}_{\rm HI}(z|\theta)}{\bar\rho_{c0}}\,,
\end{equation}
where $\bar\rho_{c0} = 3 H^2_{0}/ 8\pi G$. We show the best-fit values in \autoref{fig:OmegaHI}.
The corresponding mean HI brightness temperature
is given by \cite{Santos:2015gra}.
  \begin{equation}\label{eq:backgdeltaTbin}
  \bar{T}_{\rm HI} (z|\theta) 
  \approx 189h\frac{H_0(1+z)^2}{H(z)}\Omega_{\rm HI}(z|\theta) \,{\rm mK}\,.
\end{equation} 
With $M_{\rm{min}}$ expressed in terms of $\theta$, the dependence of the  HI bias parameters  on $R$ is shown in \autoref{fig:biasR}. 
  \begin{figure}[t!]
  \centering
\includegraphics[width=75mm,height=70mm ]{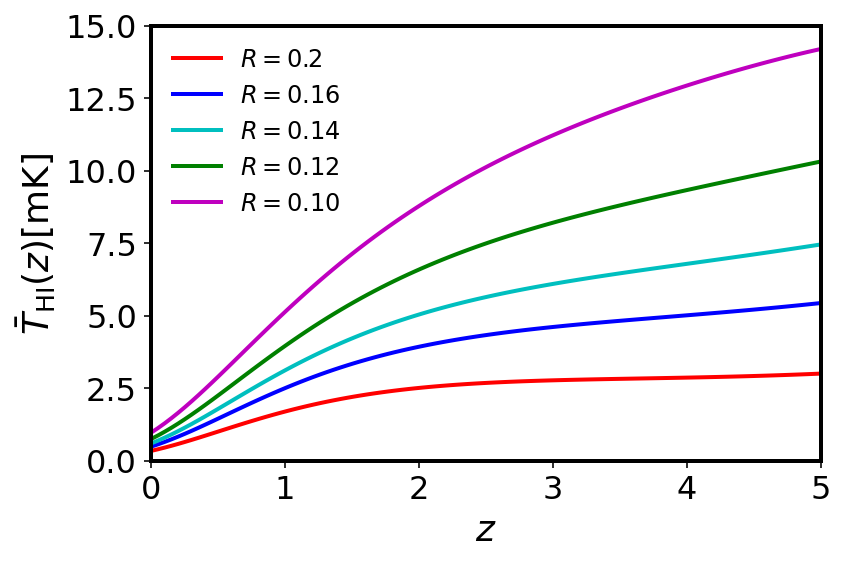}
\includegraphics[width=75mm,height=70mm ]{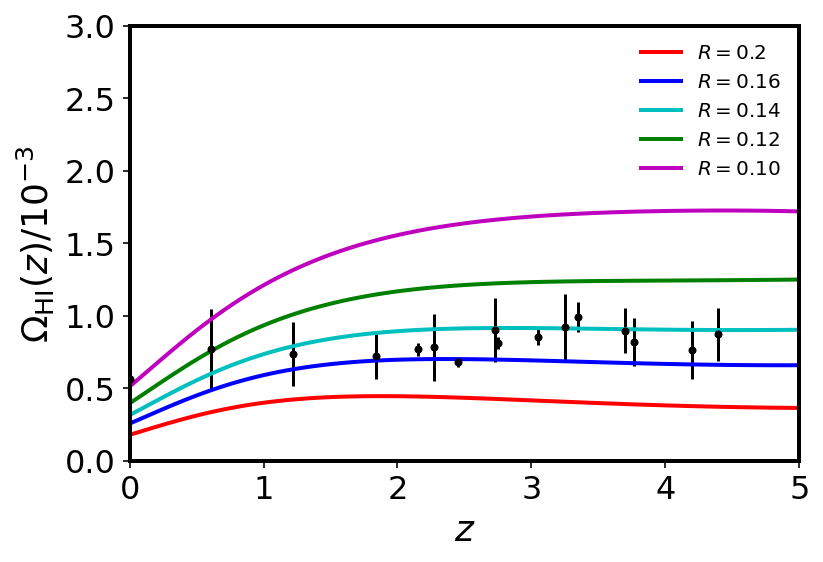}
\caption{HI brightness temperature \autoref{eq:backgdeltaTbin} (left) and  $\Omega_{\rm HI}(z)$ dependence on halo size at a fixed accretion rate $\Gamma_{1} = 2.3$, $\Gamma_{2} = -0.9$ (right). {The dark data points with error bars are observationally determined  $\Omega_{\rm HI}(z)$, they were  compiled  in~\cite{Padmanabhan:2014zma}.  $R$ is in Mpc/$h$.}
}
\label{fig:OmegaHI}
\end{figure}

\section{HI  power spectrum at one-loop and stochasticity } \label{sec:finitesize0}

We expand the smoothed HI density contrast given  in  \autoref{eq: n_m_expansion2} in Fourier space as 
\begin{eqnarray}\label{eq:delta2} 
\delta^{{R}}_{\rm HI}({\k})&=&   \mathcal{K}\one(\k) \delta_{\rm m}({\k}) 
-
 \frac{1}{2}b_2^{R}\sigma^2_{R}{\delta^{(D)}({\k})}
 \\ \nonumber &&
+ \frac{1}{2} \int\frac{\d^3 \k_1}{(2\pi)^3} \int\frac{\d^3 \k_2}{(2\pi)^3}
\mathcal{K}\two({\k}_{1},{\k}_{2}) \delta_{\rm m}({\k}_1)\delta_{\rm m}({\k}_{2})
 (2\pi)^3 \delta^{(D)} \left(  {\k}-{\k}_{1}-{\k}_{2} \right)
 \\ \nonumber &&
+\frac{1}{3!}\int\frac{\d^3 \k_1}{(2\pi)^3}\int\frac{\d^3 \k_2}{(2\pi)^3}\int\frac{\d^3 \k_3}{(2\pi)^3}
\mathcal{K}\three({\k}_{1},{\k}_{2}, {\k}_{3})
\delta_{\rm m}({\k}_1)\delta_{\rm m}({\k}_2)\delta_{\rm m}({\k}_3)
\\ \nonumber && 
\times (2\pi)^3 \delta^{(D)} \left( {\k} - {\k}_{1} - {\k}_{2}- {\k}_{3}\right)\,,
\end{eqnarray}
where  we have introduced the following Fourier space kernels
\begin{align}
\mathcal{K}\one({\k}) &= b^{R}_1 W_{R}({k})\,,
\\
\mathcal{K}\two({\k}_{1},{\k}_{2}) &= W_{R}({k}_1)W_{R}({k}_2) b_2^{R} + W_{R}({k})b_1^{R} F_2({\k}_1,{\k}_2)\,,
 \label{eq:K22term}
\\
\mathcal{K}\three({\k}_{1},{\k}_{2}, {\k}_{3}) &=W_{R}({k}_1)W_{R}({k}_2)W_{R}({k}_3)b_3^{R}+ b_1^{R}W_{R}({k})F_3({\k}_1,{\k}_2,{\k}_3)
\\ \nonumber &
 + { b_2^{R}\left[ W_{R}({k}_3)W_{R}({|{\k}_1+{\k}_2|}) F_2({\k}_1,{\k}_2)
 +2 W_{R}({k}_1)W_{R}({|{\k}_2+{\k}_3|}) F_2({\k}_2,{\k}_3) \right] }\,.
 \label{eq:K13term}
\end{align}
Here,  {we made use of the Fourier space kernels for the dark matter density field in an Einstein de Sitter universe~\cite{Bouchet:1994xp}}
\begin{equation}
{F}_2({\k}_1,{\k}_2) =  {10\over 7} + {{\k}_1\cdot {\k}_2\over  k_1 k_2}
\left({k_1\over k_2} + {k_2\over k_1}\right) + {4\over 7}\left({{\k}_1\cdot
{\k}_2\over k_1 k_2}\right)^2,
\end{equation}
and $F_3$ is given in \cite{Bernardeau:2001qr}.  {These kernels are valid as well in the $\Lambda$CDM universe provided that the $\Lambda$CDM cosmological parameters are used to evaluate the power spectrum~\cite{Scoccimarro:2000ee,McCullagh:2015oga}.}
We find that the  auto-power spectrum of the HI density contrast is given by
\begin{align}\nonumber
{P}_{\rm HI}^{(c)}({k},R)  &= {P}_{\rm HI}^{11}({k},R) + {P}_{\rm HI}^{22}({k},R)+{P}_{\rm HI}^{13}({k},R)\,, \\ 
&=\left[\mathcal{K}\one({k}) \right]^2 P^{11}_{\rm m}(k) 
+ \frac{1}{2}\int \frac{\d^3 \k_1}{(2\pi)^3}\left[\mathcal{K}\two({\k}_{1},|{\k}-{\k}_{1}|) \right]^2 P^{11}_{\rm m}({k}_{1})  P^{11}_{\rm m}(|{\k}-{\k}_{1}|) \nonumber
\\ &
\phantom=+ \frac{1}{3}\mathcal{K}\one({k})  P^{11}_{\rm m}(k) \int \frac{\d^3 \k_1}{(2\pi)^3}
\mathcal{K}\three({\k}_{1},-{\k}_{1}, {k})  P^{11}_{\rm m}({k}_{1})\,,\label{eq:PHIterm}
\end{align}
where ${P}_{\rm HI}^{11}$ is the linear HI power spectrum, $ P^{11}_{\rm m}$ is the linear matter power spectrum and ${P}_{\rm HI}^{22}({k})+{P}_{\rm HI}^{13}({k})$ constitutes  the one-loop correction.  In the long-wavelength limit, $F_2$ in $P_{\rm HI}^{22}$ vanishes.  However, the non-linear bias term  $b_{2}^R$ term {(first term in \autoref{eq:K22term})} does not vanish.
\begin{equation}\label{eq:Neff}
P_{\rm HI}^{22}(k,R) ~~ \xrightarrow{k\to0} ~~ \frac{1}{2}\left(b^{R}_{2}\right)^2  \int \frac{\d^3 \k_1}{(2\pi)^3}
W_{R}^4({k}_1)
\big({P_{\rm m}^{11}}\big)^2({k}_1) \equiv N_{\rm eff}{(R)}\,.
\end{equation}
Here, $N_{\rm eff}$, denotes the non-vanishing part of $P_{\rm HI}^{22}$ in the limit of zero momentum. This term is referred to  as induced stochasticity \cite{Baldauf:2013hka}, as white noise \cite{McDonald:2006mx}, or as the contact term \cite{Assassi:2014fva}.  It is not clear whether this term has an observational consequence \cite{Umeh:2015gza,Umeh:2016thy,Penin:2017xyf} and we revisit this issue in \autoref{sec:finitesize}. Before we proceed, we note that the same non-linear bias parameter  responsible for the re-definition of the one-point statistics in \autoref{eq:reno2} is also responsible for the emergent white noise-like feature at the two-point correlation function level on large scales. 

It is possible to analytically  simplify the terms in the last {line} of \autoref{eq:PHIterm} as 
  \begin{equation}\label{eq:HIP13renon}
 P_{\rm HI}^{13}(k,R)
 =   W^2_{R}({k}) \left\{\frac{1}{2}\left[ b^{R}_1b^{R}_3  \sigma^2_{R }+ b^{R}_1b^{R}_2\sigma^2_{b^R_1b_2^R}(k,R) \right] P_{\rm m}^{11}(k)     +{(b^{R}_1)}^2 P^{13}_{\rm m}(k)\right\}\,,
 \end{equation}
 where $\sigma^2_{b^R_1b_2^R}$ is defined below.  {$ P^{13}_{\rm m}$ is the matter power spectrum equivalent of $ P_{\rm HI}^{13}$, the full expression is given in \cite{Carlson:2009it}.}
$P^{13}_{\rm HI}$ is ultraviolet sensitive,  in  {effective field theory of large scale structure} a counter-term is usually added to remove the divergence \cite{Carrasco:2012cv,Pajer:2013jj,Konstandin:2019bay}. Here, {the problem} is parametrically controlled by the window function, {thereby eliminating the need to run an N-body simulation to calibrate the counter-term}.
Putting all this together leads to 
\begin{multline}
P_{\rm{\rm HI}}^{(c)} (k,R)= W^2_{R}({k}) \left[(b^{R}_1)^2 P_{\rm m}(k) +  \frac{1}{2} \left( b^{R}_1b^{R}_3\sigma^2_{R}+b^{R}_1b^{R}_2 \sigma^2_{b^R_1b_2^R}(k,R)\right) 
P^{11} _{\rm m}(k)\right]
\\
+ {b^{R}_{1} b^{R}_2}P_{b^R_1b_2^R}(k,R)
+ {\left(b^{R}_{2}\right)^2}P_{b^R_2b_2^R}(k,R)
+ N_{\rm eff}{(R)}
\,, \label{eq:largescalePT}
\end{multline}
where $P_{\rm m}(k) $ is the matter power spectrum up to one-loop order and 
\begin{align} 
P_{b^R_1b_2^R}(k,R) &= \frac{1}{2}\int \frac{\d^3 \k_1}{(2\pi)^3}\bigg[W_{R}({k}) W_{R}({k}_1)W_{R}({k}_2) F_2({\k}_1,{\k}_2)
\bigg]
P^{11}_{\rm m}({k}_2) P^{11}_{\rm m}(k_1)\,, \label{eq:p12}
\\   
P_{b^R_2b_2^R}(k,R) &=\frac{1}{2}\int \frac{\d^3 \k_1}{(2\pi)^3}\bigg[
W^2_{R}({k}_1)W^2_{R}({k}_2) 
P^{11}_{\rm m}({k}_2) P^{11}_{\rm m}(k_1)-W_{R}^4({k}_1)
\big\{{P_{\rm m}^{11}}({k}_1)\big\}^2\bigg]\,,\label{eq:p22}
\\  
 \sigma_{b^R_1b_2^R}(k,R)  &=
2\int \frac{\d^3 \k_1}{(2\pi)^3} \bigg[ 
  \frac{W_{R}({k}_1)W_{R}({|\k_1-\k|})}{W_{R}({k})} F_2(-{\k}_1,{\k}) \bigg] P^{11}_{\rm m}({k}_1)\,.\label{eq:sig12}
\end{align}
{ Essentially, we have decomposed $P_{\rm HI}^{22}$ term that appear in \autoref{eq:PHIterm} into scale dependent part (\autoref{eq:p22}) and scale independent part (\autoref{eq:Neff})}.
In evaluating the integrals, we  defined  $\mu_k = {\k}_1\cdot{\k}/k k_1$, and use momentum conservation ${\k}_2= {\k}-{\k}_1$, to set $k_2 = k\sqrt{r^2 - 2 r\mu_k-1} = ky$, where $k_1 = k r$, $y =\sqrt{r^2 - 2 r\mu_k+1}$.  The $k$-integrals in \autoref{eq:p12} to \autoref{eq:sig12} can be performed optimally using the FFTLog formalism~\cite{Chudaykin:2020aoj,Umeh:2020zhp}.

Finally, the HI power spectrum in  the continuous limit(\autoref{eq:largescalePT}) may be decomposed into two terms,
\begin{equation}\label{eq:continouspower}
P_{\rm{\rm HI}}^{(c)} (k,R) = P_{\rm{\rm HI}}^{(s)} (k,R) + N_{\rm eff}{(R)} \,,
\end{equation}
where $ P_{\rm{\rm HI}}^{(s)}$ is the HI power spectrum without the emergent non-linear white noise term $N_{\rm eff} $:
\begin{multline}\label{eq:signalpart}
P_{\rm{\rm HI}}^{(s)} (k,R)  =W^2_{R}({k})(b^{R}_1)^2 P_{\rm m}(k) +\frac{1}{2}\left[ b^{R}_1b^{R}_3\sigma^2_{R}+b^{R}_1b^{R}_2 \sigma^2_{b^R_1b_2^R}(k,R)\right] 
P^{11} _{\rm m}(k)
\\ \
+ {b^{R}_{1} b^{R}_2}P_{b^R_1b_2^R}(k,R)
+ {\left(b^{R}_{2}\right)^2}P_{b^R_2b_2^R}(k,R) \,.
\end{multline}
We made use of the halo-fit in CAMB to compute $P_{\rm m}$ \cite{Mead:2020vgs}, and the standard linear order Einstein-Boltzmann result from CAMB to compute $ P^{11}_{\rm m}$~\cite{Lewis:1999bs} . {The results are shown in \autoref{fig:1loop}.}
  \begin{figure}[htb!]
  \centering
\includegraphics[width=120mm,height=100mm ]{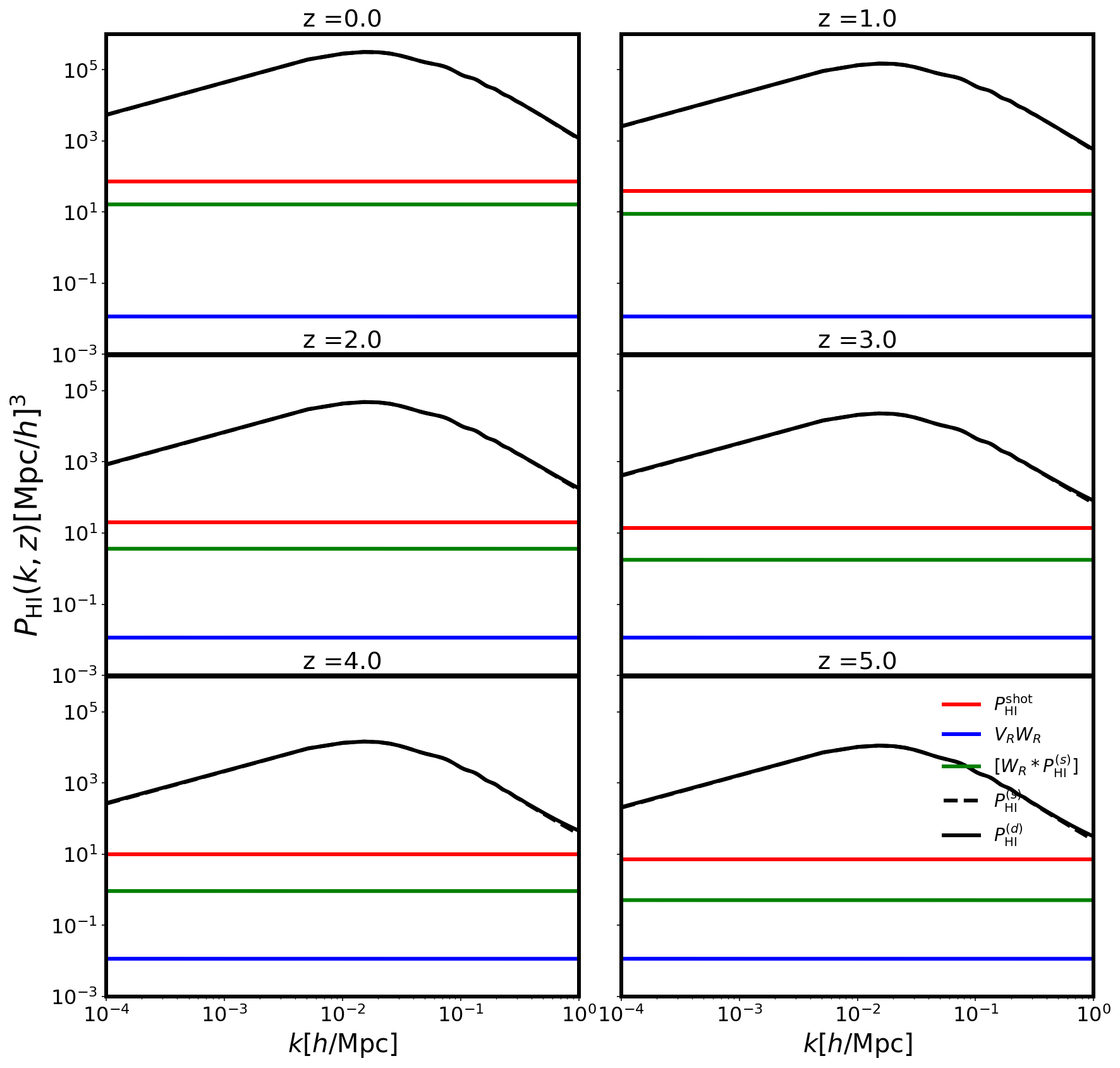}
\caption{HI power spectrum at different redshifts at the  best-fitting $R$, ${\Gamma_{1}}$ and ${\Gamma_{2}}$.
{The $k_1$-integral in \autoref{eq:fullpower4} is UV sensitive, hence, we set~$k_{\rm{max}} = 1/R$ for simplicity. }
}
\label{fig:1loop}
\end{figure}
\subsection{Finite size resolves emergent large-scale white noise problem}\label{sec:finitesize}

Here we show that $N_{\rm eff} $ {is exactly cancelled in} the full expression  of the discrete HI power spectrum when the size of haloes is taken into consideration.

Substituting \autoref{eq:continouspower} in \autoref{eq:fullpower} leads to 
\begin{equation}\label{eq:fullpower2}
   P^{(d)}_{\rm HI}(k) = P_{\rm HI}^{\text{shot}}  + N_{\rm eff}(R)- V_{{R}} W_{R}(k) - \big[{W_R}\star P_{\rm HI}^{(c)}\big](k)+P_{\rm{\rm HI}}^{(s)} (k)  \,.
\end{equation}
To appreciate the full structure of \autoref{eq:fullpower2}, we have to simplify the convolution term ${W_R}\star P_{\rm HI}^{(c)}$  further.  Using \autoref{eq:2PCFtransform} and \autoref{eq:continouspower}, with  $\x\equiv\x_{12}$,  
we obtain
\begin{align} \nonumber 
  \big[{W_R}\star P_{\rm HI}^{(c)}\big]({k}) &= \int_{{x< R}} \d ^3 \x\,\xi_{\rm HI}^{(c)} ({x})\,{\rm e}^{-i{\k}\cdot {\x}} 
={\int_{{x< R}} \d ^3 \x\big[\xi_{\rm HI}^{(s)}({x})+N_{\rm{eff}}\,\delta^{{(\rm D)}} ({\x})\big]{\rm e}^{-i{\k}\cdot {\x}}}  \,,
  \\  \nonumber  
  &=  
 { \int} \frac{\d^3 \k_1}{(2\pi)^3} \, P^{(s)}_{\rm HI}({k}_1) \,\int_{{x< R}} \d^3 \x \,{\rm e}^{i(\k_1-{\k})\cdot {\x}}
 +{N_{\rm eff}} \int_{{x< R}}\d^3 \x\,\delta^{{(\rm D)}} ({\x})\,{\rm e}^{-i{\k}\cdot {\x}}\,,
 \\ \label{eq:convolutionint} 
 &= 2V_{R}\int_0^\infty \frac{ \d k_1}{\pi} \,k_1^2\,P^{(s)}_{\rm HI}({k}_1)\,{\cal W}_R(k_1,k) +{\rm{sgn}} (R)\, {N_{\rm eff}}(R)
\,. 
\end{align}
In the last line, we used \autoref{vrwr1} to obtain the first term. 
\begin{eqnarray}\nonumber
 \int \frac{\d^3 \k_1}{(2\pi)^3}\,  P^{(s)}_{\rm HI}({k}_1) \int_{{x < R}} \d^3 \x  \,{\rm e}^{i(\k_1-{\k})\cdot {\x}} &=&
\int_{x< R} \frac{\d^3\x}{ (2\pi)^2} \,{\rm e}^{-i{\k}\cdot {\x}} 
\int_0^\infty \d k_1\,k_1^2\,P^{(s)}_{\rm HI}({k}_1)\int_{-1}^1\d\mu_1 \,{\rm e}^{i{k_1}x\mu_1} \\
&=&  2V_{R}\int_0^\infty\frac{\d k_1}{\pi}\,k_1^2\,P^{(s)}_{\rm HI}({k}_1)\,{\cal W}_R(k_1,k)\,,
\end{eqnarray}
where
\bea \label{calwr2}
{\cal W}_R(k_1,k) =
\begin{cases}
 {k\left[(k_1/k)\cos(k_1 R)\sin(kR) - \cos(kR)\sin(k_1R)\right]\over (kR)^3k_1(1-k_1^2/k^2)}~~~~\qquad {\rm{for}} ~ k\neq k_1 \,, \\ \\
{{\left[ 2k R - \sin(2 k R)\right]}\over 4(kR)^3}\hspace{5cm} {\rm{for}} ~ k= k_1 \,.  \\
 \end{cases}
\eea
For the second term, we expand the delta function in spherical coordinates  and perform the resulting integral analytically 
\begin{eqnarray}\label{eq:deltafuncint}
\int_{{x< R}} {\rm d}^3 x_{12}\delta^{({\rm D})} ({\x})e^{-i{\k}\cdot {\x}}  = \int_{0}^{R}  {\d} r \delta^{{(\rm D})}  ({r}) j_{0}(k r) = {\rm{sgn}} (R)\,,
\end{eqnarray}
where ${\rm{sgn}} (R) $ is a sigmoid function and for  $R \ge 0$, it is given by
\begin{equation}\label{eq:finitesizecondition}
{\rm{sgn}} (R) := \begin{cases}
0 & \text{if } R = 0 \\
1 & \text{if } R > 0. \end{cases}
\end{equation}
Putting \autoref{eq:convolutionint} back into \autoref{eq:fullpower2}  we obtain the final result:
\begin{eqnarray}   \label{eq:fullpower4}
P^{(d)}_{\rm HI}(k,R)   &=& P_{{\rm HI}}^{(s)} (k,R)+ P_{\rm HI}^{\text{shot}} +N_{\rm eff}(R)\left[1 -{\rm{sgn}}(R)\right]
\\ \nonumber & & \qquad  \qquad 
- V_{{R}} W_{R}(k) 
  - 2V_{R}\int_0^\infty \frac{ \d k_1}{\pi} \,k_1^2\,P^{(s)}_{\rm HI}({k}_1)\,{\cal W}_R(k_1,k)   \,.
\end{eqnarray}
\autoref{eq:fullpower4}  shows that for  $R>0$, i.e. if the finite size of haloes is taken into account, then the $N_{\rm eff}$ term  drops out exactly.
{By contrast, in the peak approximation, i.e.\ the limit where $R\to 0$ and $V_R\to0$,}  the $N_{\rm eff}$ term does {\em not} drop out,  so we recover the standard result \cite{McDonald:2006mx,McDonald:2009dh,Assassi:2014fva,Umeh:2015gza,Umeh:2016thy,Penin:2017xyf}
\begin{equation}\label{eq:fullpower5}
   P^{(d)}_{\rm HI}(k) = P_{\rm{\rm HI}}^{(s)} (k)+ P_{\rm HI}^{\text{shot}}  + N_{\rm eff}
   \,.
\end{equation}
  \begin{figure}[h!]
  \centering
\includegraphics[width=75mm,height=70mm ]{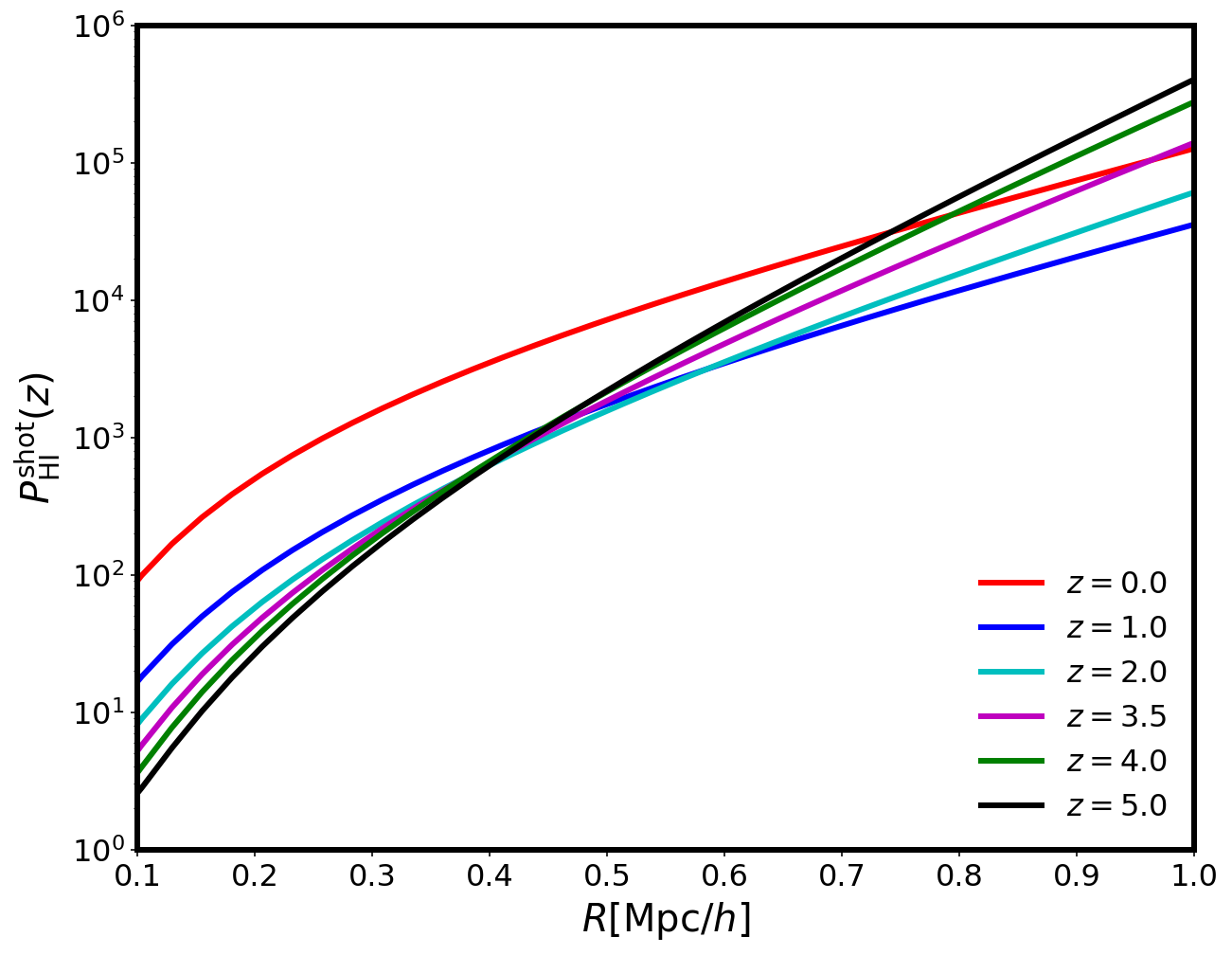}
\includegraphics[width=75mm,height=70mm ]{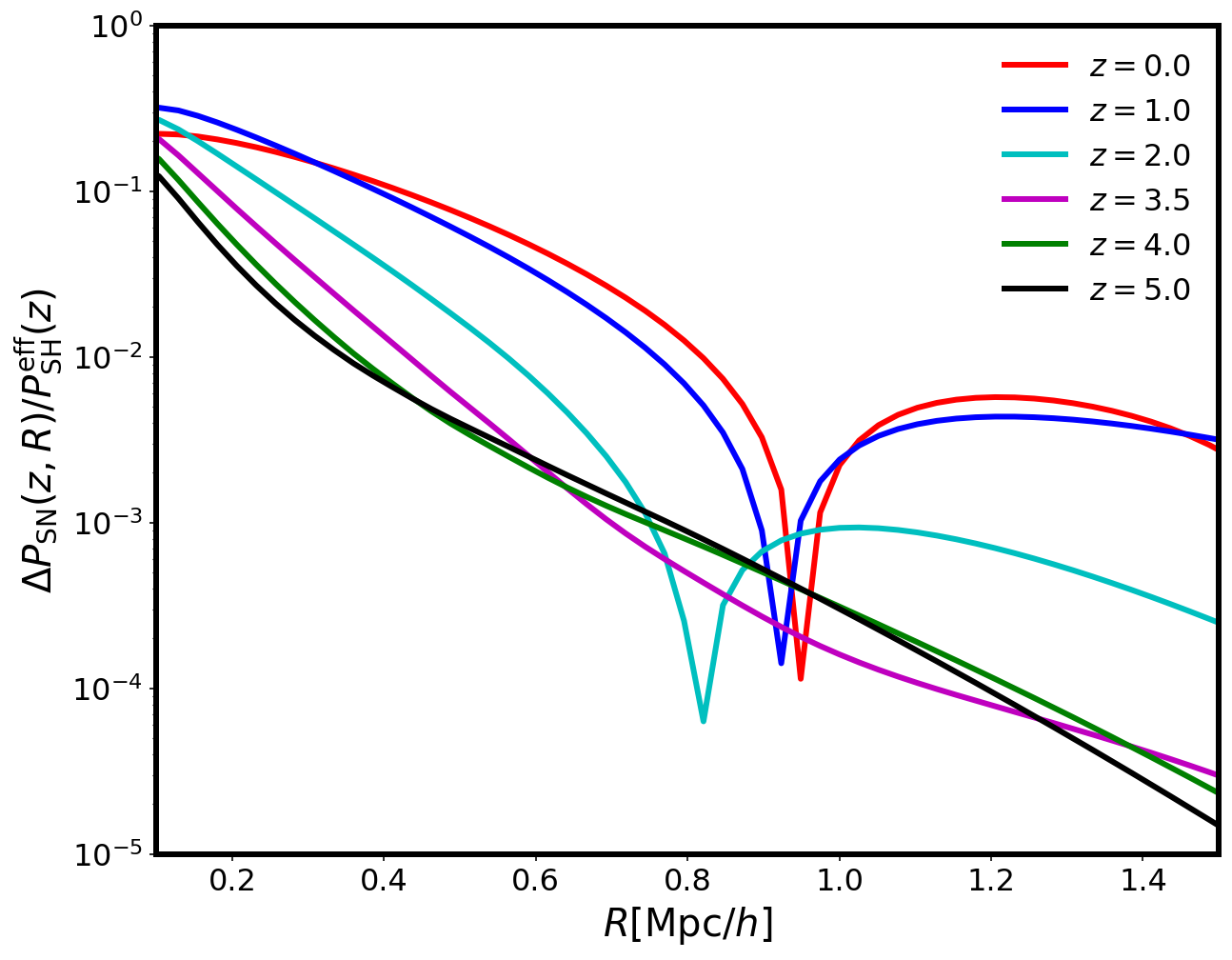}
\caption{HI shot noise at the best-fit value of accretion rate(left). {Difference  between effective shot noise and intrinsic shot noise at the best-fit value of accretion rate(right).  For the effective shot noise we set $k = 0.001$~h/{\rm{Mpc}}.}
}
\label{fig:shot}
\end{figure}

\autoref{eq:fullpower4} is an important result. We have shown, for the first time, that the emergent large-scale white noise is as a result of the breakdown of the peak approximation for the tracer density field.  This provides more clarity on how to handle $N_{\rm eff}$ in standard perturbation theory or in effective field theory approaches -- avoiding the need to  set $N_{\rm eff}$ to zero~\cite{Assassi:2014fva} by hand or absorbing it into the shot noise~\cite{McDonald:2006mx,McDonald:2009dh} which does not work for the HI brightness temperature~\cite{Umeh:2015gza,Umeh:2016thy,Penin:2017xyf} since the halo model predicts shot noise  with  smaller amplitude when  compared to the amplitude of $N_{\rm eff}$. We have now shown that this term vanishes exactly as soon as the finite size of the source is taken into account. 

Comparing \autoref{eq:fullpower4} to  \autoref{eq:Peeblesapproximation}, the second and third terms can be re-written as part of the  effective shot noise contribution
\begin{align}\label{eq:effectiveshotnoise}
P^{\rm{eff}}_{\rm{SN}}(k,R) &= P_{\rm HI}^{\text{shot}}  - V_{{R}} W_{R}(k) -  2V_{R}\int_0^\infty\frac{\d k_1}{\pi}\,k_1^2\,P^{(s)}_{\rm HI}({k}_1)\,{\cal W}_R(k_1,k)\,,
\\
&\approx  P_{\rm HI}^{\text{shot}}  \qquad {\rm{for}}  \qquad R <1~{\rm{Mpc}/h}\,,
\end{align}
where the Poisson shot noise, $P_{\rm HI}^{\text{shot}}, $ is obtained by weighting the halo density field appropriately with  the HI-halo mass relation \cite{Bull:2014rha, Padmanabhan:2016fgy}, namely
\begin{equation}\label{eq:HIshotnoise}
P_{\rm HI}^{\text{shot}}(R) =\frac{1}{\bar{n}^{R}_{\rm{HI}}} =  \frac{1}{{\bar\rho}^2_{\rm HI} (z)}\int_{M_{\rm{min}}}^{\infty} \d M \, M_{\rm HI}^2(M)\,
n_h(z,M)\,.
\end{equation}
For $R \ll 1  \, {\rm Mpc}/h$, the second term  and the  third term are negligible,  see \autoref{fig:shot}. For  massive halos, i.e $R \ge 1  \, {\rm Mpc}/h$, one would expect the amplitude of the effective shot noise term to be substantially modulated by the size of the excluded region, as inthe right panel of  \autoref{fig:shot}.
%
%
{In the right panel of \autoref{fig:shot}, we plot the fractional difference between the effective shot noise and the intrinsic shot noise: $\Delta P_{\rm{SN}}(k,R)  = P^{\rm{eff}}_{\rm{SN}}(k,R) - P_{\rm HI}^{\text{shot}}$.  This shows that for HI with $R > 0.14  \, {\rm Mpc}/h$, the effect of the sub-Poissonian  noise correction peaks at the comoving boundary of the halo containing HI, it decreases away from the halo boundary.  We find that increasing $R$ while keeping every other parameter fixed for the HI, the contribution from  the last term in \autoref{eq:fullpower4} changes sign at  $R $ away from the comoving halo boundary.  }


\section{Summary and Outlook}\label{sec:Discussionandconc}

\subsection{Consequences of the exclusion region}

The standard halo model framework and its various extensions \cite{Hadzhiyska:2019xnf,Hadzhiyska:2020tec} split the mass distribution in the Universe into distinct regions.  The correlation function is split into correlations between the distinct regions and correlations within each region. These are the well-known one- and two-halo terms for the two-point correlation function. 
Standard perturbation theory, on the other hand, does not make this distinction, it assumes that the perturbation theory expansion is valid at all locations, even within highly dense  dark matter haloes.  This assumption leads to the well-known ultraviolet problems at the loop level and non-linear white noise in the infra-red due to contact terms \cite{Assassi:2014fva}. 

We have argued that the physics of clustering  imposes a physical scale which is related to the dynamically defined splashback radius of haloes that allow treatment of structure evolution in line with the halo model. 
There are few key points to highlight:
\begin{itemize}
\item {\tt{Significance of the splashback radius:}}  {We have proposed a connection between the minimum halo mass  that can host HI and  the physical halo boundary.  This connection involves the mass accretion rate,  which  impacts, the growth of the physical halo boundary. The  accretion rate and halo boundary are physical parameters that  future surveys such as  HI  intensity mapping  with HIRAX~\cite{Newburgh:2016mwi},  MeerKAT~\cite{Santos:2017qgq}, SKA~\cite{Santos:2015gra}  and others  could provide an opportunity to constrain.  }

\item {\tt{Cancellation of the induced white noise:}} {Any tracer power spectrum at one-loop within the standard perturbation theory has a component  which does not vanish in the limit of zero momentum. This behaviour is also present in the standard halo model of the matter power spectrum, in this case, the one-halo term leads to a non-zero contribution in the limit of zero momentum~\cite{Chen:2019wik}.  The non-vanishing component is referred to  as induced stochasticity in \cite{Baldauf:2013hka}, as white noise in \cite{McDonald:2006mx}, or as the contact term that leads to a delta function in real space in \cite{Assassi:2014fva}.  It was observed in \cite{Umeh:2015gza,Umeh:2016thy,Penin:2017xyf} that it could have consequences for the bias parameters on large scales for the HI brightness temperature if indeed  it is physical. We have shown that this term vanishes exactly as soon as the size of the discrete source is taken into account. }

\item {\tt{Mass weighting of haloes}:} {We have shown that it is possible to minimise  $P^{\rm{eff}}_{\rm{SN}}(k,R) $  by  taking the finite halo size into consideration. Setting $R$ to its corresponding physical boundary value gives the minimum effective shot noise (see \autoref{fig:shot}).   A similar idea has been explored in estimating the halo power spectrum from the N-body simulations but in that context,  it is known as mass weighting of haloes~\cite{Seljak:2009af}.   The connection between mass weighting of haloes and halo exclusion criteria  was made in \cite{Baldauf:2013hka}. Essentially, choosing different mass bins(mass weighting)  corresponds to choosing  the most optimal $R$  that minimaxes $P^{\rm{eff}}_{\rm{SN}}(k,R) $  the most. 
%
This idea  was used in \cite{Hamaus:2010im,Hamaus:2012ap} to show how shot noise associated with the halo power spectrum could be significantly suppressed  on large scales thereby improving the signal to noise ratio.    }

\item  {\tt{Sub-Poissonian process}:} {We have shown that the noise associated with discrete tracers  is not  entirely due to a Poisson process, i.e. $P^{\rm{eff}}_{\rm{SN}}(k,R) \neq 1/\bar{n}_{\rm HI}$, there would be a sub-Poissonian contribution.  This feature has already been observed both  in cluster auto-and cluster-galaxy cross-correlations of the Sloan Digital Sky Survey \cite{Paech:2016hod}.}


\end{itemize}


\subsection{Conclusions}\label{sec:conc}

Our understanding of the universe through large scale structure has relied heavily on the halo model \cite{Cooray:2002dia}. Haloes in this context are not point sources, but rather  virialized extended objects with finite boundaries given by the splashback radius \cite{Adhikari:2014lna}.  Luminous objects such as galaxies and neutral hydrogen are situated inside haloes and held  together by the  self-gravitational field of haloes \cite{Sheth:1998xe}.  However for the halo power spectrum in standard perturbation theory, haloes are modelled as point sources on all scales \cite{Perko:2016puo}.  We have  shown how to take into account the physical constraints  due to the finite size of the halo in modelling  the power spectrum  of the HI brightness temperature.  

The standard practice is to model tracers as point sources, in this limit the power spectrum is given as a sum of the continuous power spectrum and  the Poisson  shot noise (see \autoref{eq:Peeblesapproximation}). The Poisson shot noise is given by the mean number density of the tracer. In addition to the Poisson shot noise, there is also a non-linear white noise contribution from the continuous power spectrum in the limit of zero momentum. This suggests a break down of the mass-energy conservation for tracers \cite{Chen:2019wik}. We have shown that taking into account the finite size of haloes  introduces two additional terms to the power spectrum of discrete tracers (see \autoref{eq:fullpower}).  The two  extra terms describe  the fact that information with wavelength less than the size of  haloes is uncorrelated and therefore defines the exclusion region. We have shown that taking the exclusion region into account leads to an \textit{exact cancellation} of the non-linear white noise-like term that appears in the limit of zero momentum (see \autoref{eq:fullpower4}).  We showed that the effective shot noise contribution on large scales  is sub-Poissonian  and it is dependent on the size of the exclusion region.

For the HI brightness temperature, we argued that the exclusion region  is naturally given by the splashback radius of the halo  with the minimum mass required to host HI.  We describe how the HI brightness temperature within haloes  of a given mass or size  may be modelled as a tracer of the dark matter density  field.  Finally, we argued that taking into account the consequences of the finite size of haloes in modelling the power spectrum of any tracer may explain why the concept of mass-weighting of haloes  improves the signal-to-noise ratio \cite{,Hamaus:2010im,Hamaus:2012ap}.





\section*{Acknowledgement}

We would like to thank 
 Guido D'Amico, Benedikt Diemer and Kazuya Koyama for useful discussions.
OU  is supported by the  UK Science \& Technology Facilities Council (STFC) Consolidated Grants Grant ST/S000550/1. RM is supported by  the South African Radio Astronomy Observatory (SARAO) and the National Research Foundation (Grant No. 75415), and by the UK STFC Consolidated Grant ST/S000550/1.
HP acknowledges support from the Swiss National Science Foundation under Ambizione Grant PZ00P2\_179934. SC also acknowledges the support from the Ministero degli Affari Esteri della Cooperazione Internazionale - Direzione Generale per la Promozione del Sistema Paese Progetto di Grande Rilevanza ZA18GR02. 
We made use of emcee \cite{Emcee:2019JOSS....4.1864F} and zeus-mcmc \cite{Karamanis:2020zss}  for the statistical analysis and getdist \cite{Lewis:2019xzd} for visualisation.  Also, we used  xPand~\cite{Pitrou:2013hga} for perturbation theory expansion.


\appendix

\section{Basic tools of the halo model}\label{sec:halomodel}

We calculate the bias parameters from the Sheth-Tormen halo mass function for a spherical collapse model \cite{Sheth:1999mn}: 
\begin{equation}\label{eq:sheth-tormen}
n_h(M) = \nu f(\nu) \frac{\bar{\rho}}{M^2}\frac{\d\ln \nu}{\d\ln M},
\end{equation}
where the peak height $\nu$ is related to the variance in dark matter density field, $\sigma^2_{\rm m}$, $\nu = (\delta_c/\sigma_{\rm m})^2$ and $\delta_c = 1.686$ is  the critical threshold for a spherical collapse at the current epoch  obtained from linear perturbation theory.  A halo of  mass $M = \bar{\rho} V$  is formed when the walk first crosses a barrier $f(\nu)$:
\begin{equation}
\nu f(\nu)=A(p)\left(1+\frac{1}{(q
\nu)^p}\right)\sqrt{\frac{q\nu}{2\pi}} \exp\left(-\frac{q\nu}{2}\right),
\end{equation}
where $q=0.707$ and $p=0.3$ are obtained from a fit to numerical simulations.  The halo bias parameters up to third order are given by \cite{Cooray:2002dia}
\begin{align}\label{eq:Multibias1}
b^{h}_1   &=1+\frac{(q\nu-1)}{\delta_c}+\frac{2 p}{\delta_c\left(1+(q\nu)^p\right)}\,,\\
b^{h}_2   &=\frac{8}{21}\left(b_1   -1\right)+\frac{4\left(p^2+ \nu p q\right)-(q\nu-1)\left(1+(q\nu)^p\right)-2p}
{\delta^2_c \left(1+(q\nu)^p\right)}
 +\frac{1}{\delta_c^2}\left((q\nu)^2 - 2q\nu -1\right)\,
\label{eq:Multibias2}\\
b^{h}_{3} &= -\frac{236}{189}\left( b_{1}-1\right) -\frac{13}{7} \left(b_{2}- \frac{8}{21}\left( b_{1}-1\right)\right)
-\frac{\left(3 + 3 \nu q + 3 \nu^2 q^2 - \nu^3 q^3\right)}{\delta_c^3}
\\ \nonumber &
+ \frac{\left(8 p^3 + 12 p^2 \left( 1 +\nu q\right) + p \left( 6 \nu^2 q^2 -2 \right) \right) }{\delta_c^3 \left( 1 + 1+ (\nu q)^p\right)} 
+6\frac{\left(1+ 2\nu q -\nu^2 q^2\right)}{\delta_c^3} - 24 \frac{\left( p^2 + \nu pq\right)}{\delta_c^3\left( 1+ (\nu q)^p\right)}
\\ \nonumber &
 -4\frac{(1-\nu q)}{ \delta_c^3} + 8 \frac{p}{\delta_c^3\left( 1+ (q\nu)^p\right)}\,,
\label{eq:Multibias3}
\end{align}

\def\aj{AJ}                   
\def\araa{ARA\&A}             
\def\apj{ApJ}                 
\def\apjl{ApJ}                
\def\apjs{ApJS}               
\def\ao{Appl.Optics}          
\def\apss{Ap\&SS}             
\def\aap{A\&A}                
\def\aapr{A\&A~Rev.}          
\def\aaps{A\&AS}              
\def\azh{AZh}                 
\def\baas{BAAS}
\def\jcap{JCAP}
\def\jrasc{JRASC}             
\def\memras{MmRAS}
\def\na{New Astronomy}
\def\nat{Nature}
\def\mnras{MNRAS}             
\def\pra{Phys.Rev.A}          
\def\prb{Phys.Rev.B}          
\def\prc{Phys.Rev.C}          
\def\prd{Phys.Rev.D}          
\def\prl{Phys.Rev.Lett}       
\def\pasp{PASP}               
\def\pasj{PASJ}
\def\physrep{Phys. Repts.}
\def\qjras{QJRAS}             
\def\skytel{S\&T}             
\def\solphys{Solar~Phys.}     
\def\sovast{Soviet~Ast.}      
\def\ssr{Space~Sci.Rev.}      
\def\zap{ZAp}                 
\let\astap=\aap
\let\apjlett=\apjl
\let\apjsupp=\apjs

\providecommand{\href}[2]{#2}\begingroup\raggedright\endgroup

\end{document}